# Descriptive analysis of computational methods for automating mammograms with practical applications

A. Bhale[a], M. Joshi[b]

[a,] Transcend Research Consultancy, Pune, Maharashtra, India

[b]School of Computer Sciences, North Maharashtra University,

Jalgaon, Maharashtra, India

**Abstract** Mammography is a vital screening technique for early revealing and identification of breast cancer in order to assist to decrease mortality rate. Practical applications of mammograms are not limited to breast cancer revealing, identification ,but include task based lens design, image compression, image classification, content based image retrieval and a host of others. Mammography computational analysis methods are a useful tool for specialists to reveal hidden features and extract significant information in mammograms. Digital mammograms are mammography images available along with the conventional screen-film mammography to make automation of mammograms easier. In this paper, we descriptively discuss computational advancement in digital mammograms to serve as a compass for research and practice in the domain of computational mammography and related fields. The discussion focuses on research aiming at a variety of applications and automations of mammograms. It covers different perspectives on image preprocessing, feature extraction, application of mammograms, screen-film mammogram, digital mammogram and development of benchmark corpora for experimenting with digital mammograms.

**1. Introduction**

Breast cancer is a kind of cancer with the maximum occurrence rates in women. It is one of the the major reasons of malignancy related casualty cases in women across many countries of the world. In 2020, an projected 276,480 new cases of invasive breast cancer are anticipated to be identified in women in the U.S., along with 48,530 new cases of non-invasive (in situ) breast cancer [17, 79]. The Indian Council for Medical Research recently issued a report which stated that in 2016 the total number of new cancer cases is estimated to be about 14.5 lacks. This number will probably increase to 17.3 lacks in 2020 [56].

Methods that can accurately predict breast cancer are greatly needed. Mammography is one of the most efficient methods for early detection of breast cancer. It is a definite type of breast imaging system that utilizes small amount of x-rays or magnetic resonance imaging to detect cancer early. However, it is difficult to manually analyse mammograms because of the deficiency in precision, two dimensional views and complexities normally linked with images.

Lots of researchers are trying to enhance mammography with the help of computational or algorithmic methods. Many researchers are operational in diverse methods to compose mammograms more significant. A high quality image would be more useful to improve the accuracy of mammogram investigation. But, conventional mammograms are formed on screen-film and a expert has to examine the images on a light box. This equipment has been used in identifying breast cancer, but it's accuracy is less because of its low accuracy score.

The intrinsic limitations of traditional mammogram analysis methods are surmounted by the emerging innovation of digital mammography. The necessity of unambiguous results in digital mammograms is possible because a perfect combination of sensitivity and specificity has not been achieved. This leaves room for further improvement in the available technologies for mammogram analysis.

Hence, researchers are striving to improve mammogram investigation using computational influence of modern computer systems. Mammograms are processed using diverse computational techniques to achieve enhanced images for the purpose of effective analysis. The objective of this paper is to discuss various computational methods employed for successful mammogram analysis. This would serve as a directional compass for research and practice in the domain of computational mammography and related fields such as computer vision, medical image analysis, computer graphics and image processing. The discussion focuses on research that are aimed at a variety of mammogram automations and practical applications of mammograms. Specifically, it covers different perspectives on image preprocessing, feature extraction, application of mammograms, screen-film mammogram, digital mammogram and development of standard corpora for experimenting with digital mammograms. Figure 1 presents a succinct summarisation of these categories of different computational methods for analysis of mammograms as discussed in this paper.

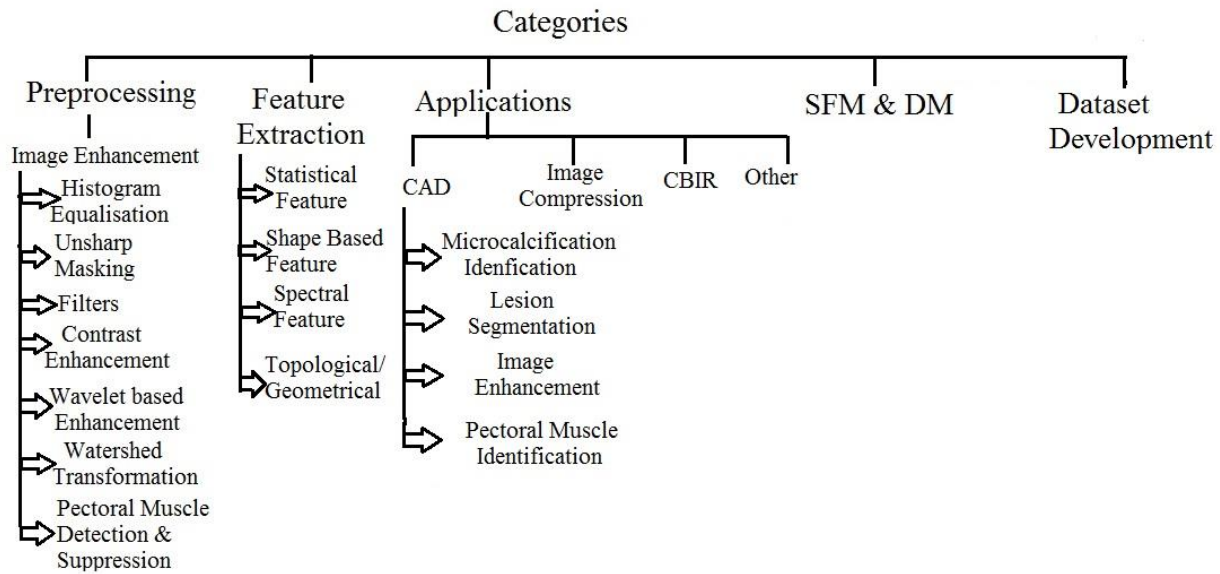

Figure1 Computation methods for analysing and automating mammograms.

Briefly, image preprocessing plays a vital role in organizing suitable input for the succeeding phase of Computer Aided Diagnosis (CAD). The accomplishment of CAD mainly depends upon how well mammograms are enhanced. Moreover, feature extraction plays an important role in object recognition and classification systems because it describes relevant properties and information contained in an object. Feature extraction impacts on the performance of object recognition and classification systems. The absence of effective feature extraction methods would imply the challenges of curse of dimensionality that is relevant to numerous learning algorithms and it denotes the severe augment in computational intricacy to rise in categorization inaccuracy [76].

The importance of breast cancer problem to clinical practice provides sufficient momentum to formulate application of digital mammography a medical truth. Taking an inventory of different applications of mammograms cannot be over emphasised because mammography offers promise to improve image quality and interpretation. Consequently, numerous study functions that are related with mammograms are discussed. In addition, the views of researchers on the use of screen-film and digital mammograms are discussed. The development of mammogram corpora is discussed in this paper because the reliability of database of mammograms can play a significant task in the development of mammography research by providing researchers with authentic benchmark mammogram data to support research experiments.

## 2. Image Preprocessing

Raw mammography representations are not directly appropriate for medicinal prediction because of the presence of artefacts. The images worth a preprocessing stage in order to disclose unambiguous characteristics that can help during a decisive decision making. Image enhancement is an important aspect of a preprocessing phase that helps to suppress the effects of artefacts. Different computational methods for the enhancement of mammograms that are categorised into seven distinct categories are discussed in this section.

### 2.1. *Histogram Equalization with its Variants*

Sivaramakrishna et al. [1] experimented with a improved version of novel Histogram Equalization algorithm. Histogram equalization is a broadly utilized technique to augment disparity in an image. The method has been developed to improve its capability and one of its famous variants is the contrast limited adaptive histogram equalization (CLAHE) algorithm recommened by Pizer et al. [2]. Experimentation was performed using CLAHE to enhance the performance of mammograms by correcting the problem of image over enhancement [senthilkumar 3]. Subsequent to testing with four diverse image enhancement algorithms, the authors suggested that CLAHE technique significantly transforms the appearance of images as compared to further methods and is less chosen by radiologists.

An edge detection algorithm was proposed that uses CLAHE as an enhancement method to preprocess mammograms [3]. When the proposed algorithm was applied on mammograms preprocessed using CLAHE, tumors emerge noticeable in the backdrop and more features of images were confined. Rahamati et al. [4] further improved CLAHE algorithm by adding a nonlinear fuzzy function that overcomes the lacuna experimented with CLAHE. Graininess was introduced because of the enhancement of noisy images when CLAHE method was used. The algorithm not only eradicates noise and concentration in homogeneities in the backdrop, it also maintains the normal gray level deviations of mammography images inside doubtful lesions. The results obtained on real mammography images showed improved segmentation accuracy as compared to the non preprocessed images.

Sundaram et al. [5] initiated a new disparity in histogram equalization, which they termed Histogram Modified Local Contrast Enhancement (HM-LCE). The method regulates the stage of contrast enhancement ensuing into a strong contrast image and emphasizes the local features present in the original image for more pertinent analysis. The accuracy of this technique was establishd using the images with microcalcification gained from the Mammography Image Analysis Society (MIAS). Jasmine et al. [6] utilized the adaptive histogram equalization to get better contrast in images. In this customized technique, numerous histograms were acquired, all equivalent to a different fragment of an image.

All of these histogram based image enhancement methods reallocate the weightlessness significances of the image. Abraham et al. [7] reviewed methods like thresholding, region based segmentation, fuzzy c-means clustering segmentation, k-means clustering, watershed and edge detection. They discussed commonly used mammogram image segmentation methods for early revealing and identification of breast malignancy.

## 2.2 *Unsharp Masking with its Variants*

Sivaramakrishna et al. [1] used experiments to find out with the aim of mammograms having mass deposition, radiologists preferred pictures preprocessed using the adaptive unsharp masking over other image preprocessing algorithms. The adaptive unsharp masking algorithm proposed by Ji et al. [8] presents appropriate increase feature that differes for every pixel. It is based on limited neighbourhoods of every pixel in the original image. Panetta et al. [9] presented a Non Linear Unsharp Masking algorithm (NLUM) for mammogram enhancement. The technique presents suppleness in permiting manual assortment of NLUM factor and dissimilar types of filters can be embedded into the nonlinear sorting operator. Furthermore, the improved filtered part of the image can be gained by the combination of diverse linear or nonlinear procedures. The preprocessing has resulted in the enhancement of mammogram and better performance of illness identification. A collection of mammography images was improved using four unlike image enhancement algorithms including NLUM. The enhanced images were provided to human radiologists to find their subjective assessments and NLUM enhanced images were preferred by the radiologists who contributed in the study.

## 2.3. Image *Filter*

Image filter has important application in mammogram analysis for suppressing artefacts and segmenting region of interest. Image filters such as Smallest Unvalued Segment Assimilating Nucleus (SUSAN) have been used to extract areas such as micro-calcifications in mammograms that have got specific intensity brightness and size. The 2-d median filter and thresholding methods have been used for noise removal in the mammogram applications. Liu et al. [12] utilized the median filter method to remove noise in mammograms. Paradkar et al. [10] applied plain thresholding pursueed by Neuron-based thresholding to sort out figure pixels that are not applicants for microcalcification. The picture is clustered, in which the cluster centroids were regarded as the seed pixels. The early seed points for clustering were gained with a neuron-based procedure. The cluster centroids were utilized to decide the second level threshold. In their trial more than one third of the original number of picture pixels was sorted out. The consequential image was further used in the automatic microcalification recognition procedure.

Jasmine et al. [6] applied global gray level thresholding to preserve the pixels between a pre-selected upper-threshold and lower-thresholding of the gray level histogram. This preprocessing was pursued by counterlet investigation for feature mining. The mammograms were automatically classified using the Artificial neural Network (ANN) pattern classification method. Nagi et al. [11] recommended a technique for detection of Region of Interest (ROI) in an image and to facilitate that they applied various preprocessing methods of image enhancement.

Dina Abdelhafiz et al. [60] recommend to use a U-Net method to automatically identify and fragment lesions in mammogram images. Authors mentioned that U-Net [61], which is an end-to-end convolutional neural network (CNN) based model that has achieved remarkable results in segmenting bio-medical images [62]. They proposed a model and compared it with R-CNN and Region Growing models. Authors concluded that the performance of the proposed Deep Learning model show potentials to make its functional application probable for medical applications to support radiologists.

**2.4.** *Contrast Enhancement with it Variants*

The adaptive neighbourhood contrast enhancement algorithm was mostly the desirered option amongst four enhancement algorithms estimated by Sivaramakrishna et al. [1] for microcalcifications. The adaptive neighbourhood contrast enhancement algorithm put forwarded by Morrow et al. [13] applies seed pixels to outline a local region by adding neighbouring pixels inside a particular gray level divergence. Such local areas are called the foreground, which is used to formulate background. Morrow et al.[13] propositiond an experimentally resultant look-up chart to verify the value of contrast. The process was found better for mammogram images presenting microcalcification than for images showing masses.

Nordin et al. [14] put forwarded a alteration in a straightforward contrast enhancement process by generating several diverse images equivalent to a unlike variety of intensity levels. These descriptions were exercised to outline a moving video to exhibit evolution from one intensity level to another. Sample mammograms were enhanced using the proposed method and frames at different intensity levels were displayed. The authors concluded that instead of using complex variations of contrast enhancement algorithms, the proposed modifications in a simple contrast enhancement algorithm can also provide significant results. Potente et al. [77] reported on the practical utilization of dynamic contrast enhanced magnetic resonance mammography to diagnose breast lesions. They followed an interpretive model that was based on a diagnostic method, which combined contrast enhancement with morphological analysis. Rangayyan et al. [15] presented a evaluation of CAD of breast disease and analysed research work that used image contrast enhancement methods.

Sara Dehghani et al. [72] proposed a method to improve surroundings as mammography images have a gloomy surroundings, which is not significant in mammogram processing. Three stages were applied to obtain the enhancement. In the first step, excessive image parts that are on two sides of the image are omitted using pixel brightness. In the second step, the dissimilarity of the breast direction is completed using a threshold boundary of gray level of the two bisects of the image and put all images in one direction. The third step is the breast section segmentation from the surroundings. They do this work by means of a sequence of point functions and region growing method.

### 2.5. *Wavelet based Enhancement*

The wavelet based multiresolution look likes a multiresolution technique that presents in the human vision structure. A picture can be measured as a set of pixels including primal characteristics at unlike scales. A wavelet disintegration of an image roughly splits the image into numerous subbands. In the case of mammogram, tiny elements like microcalcifications will be important in one subband, whereas bigger features like masses will be leading in a unusual subband. A number of methods of wavelet based enhancements have been recommended. Sersic et al. [16] utilized redundant separate wavelet make over to detect microcalcification, in which the Bspline wavelet was implemented to gain five diverse images. In their multiresolution disintegration of mammograms, the first level detail coefficients (HH, HL and LH) typically include noise. Detail coefficients in the levels 2 to 5 hold superior breast structure and microcalcifications with some noise. The level 5 estimatation coefficients (LL) include low occurrence background that corresponds to the tissue compactness. Sub images were improved and used to rebuild the ultimate image. The authors stated that their method showed good results.

The dyadic wavelet enhancement algorithm was applied in contrast with other algorithms by Sivaramakrishna et al. [1]. The input image was first crumbled into a set of subband images using suitable scrutiny filters. The image can be recreated or produced from its subband images using amalgamation filters. It was experimented that for images that show microcalification, wavelet based augmentation was preferred to adaptive unsharp masking and contrast-limited adaptive histogram equalization enhancement algorithms. Sakka et al. [18] utilized the Haar, Daubechies, Coiflets and least asymmetric Daubechies wavelet based enhancements. A proportional study amongst the orthogonal wavelets was offered in order to decide on the suitable family to be utilized for mammography image processing. Investigational examination with the mammography images gaind using the Matlab wavelet toolbox were presented and compared.

## 2.6. *Watershed Transformation*

The watershed transformation is a region-based segmentation method widely used in mathematical morphology and medical image segmentation. In general, the watershed transformation is simple and can be worked out on the gradient illustration so that the margins are find at high gradient points. It can be utilized for gray scale images, textural images and binary images in three dissimilar ways, which are distance transforms, gradient and marker controlled. Xu et al. [19] suggested the marker-controlled watershed for lesion segmentation in mammograms. The marker-controlled watershed transformation was futured to prevail over the occurrence of various local minima and in excess of segmentation in the watershed method. An original mammogram was converted into the morphological gradient image by implementing morphological dilation and erosion operators. Morphological reform consequences in a soft morphological gradient image that had been utilized for watershed analysis and for dermination of lesion boundry.

In addition, Camilus et al. [20] made use of the watershed transformation for pectoral muscle detection in a mammogram. Even though they initiated with similar steps as that of Xu et al. [19] to gain a evened gradient image. They recoomended a amalgamation algorithm to defeat the trouble of over segmentation allied with the watershed analysis. The outcome of automatic detection of pectoral muscle boundary were authenticated by evaluating with manually recognized pectoral muscle boundaries. Ponraj et al. [22] conducted a survey of preprocessing techniques for mammograms, illustrated the watershed transformation and reported in a few studies that they have utilized the watershed transformation. The watershed segmentation was used in an automated method to help radiologists in detecting Micro Calcification Clusters(MCC) in digital mammography [23]. In the paper, the authors studied various methods such as wavelet transformation, morphology operation, adaptive threshold and watershed segmentation. They ended by quoting that watershed segmentation method is extra precise than the adaptive thresholding technique, but it requires additional time.

### 2.7. *Pectoral Muscle Detection and Suppression*

The procedure of recognizing pectoral muscle itself requires removal of noise from mammograms. The detection and elimination of pectoral muscle can be regarded as preprocessing for CAD. We have studied a few reasrch papers that detailed on the techniques used to make out and eradicate pectoral muscle from mammogram for added processing. Nagi et al. [11] emphasised the necessitate to split the breast contour section by spoting and restraining the pectoral muscle section. They declare that the occurrence of pectoral muscle in mammograms prejudicees additional automated revealing method. The authors projected morphological preprocessing pursued by seeded region growing algorithm to eventually shape out breast profile section by eliminating the pectoral muscle from the mammogram. The algorithm was applied using mammograms of contradictory densities from several databases and precisely matches with manual assessment by a radiologist.

Bandyopadhyay et al. [24] implemented preprocessing to mammogram and suggested a effortless technique to sort out a pectoral muscle section. The authors asserted that a reversed right angle triangle formed at the top left angle of a mammogram wraps the whole section of pectoral muscle. Moreover, they declared that this procedure consequences in a complete accomplishment ratio of 80 percent on diverse pairs of mammogram of dissimilar profiles and dimensions. Camilus et al. [20] utilized the watershed transformation technique to distinguish the pectoral muscle boundary. A exclusive watershed line matching to the pectoral muscle rim emerges when a watershed conversion was implemented to a mammogram.

The inherent crisis of over segmentation grounded by the watershed transformation process was overcomed by scheming an efficient algorithm that combined sub basins in the pectoral muscle section to outline a solitary amalgamated region of interest [25]. A relative outcome of diverse computational methods was mentioned using false positive and false negative standards for the evaluation. A relative investigation of Fuzzy C-Means (FCM), K-Means (KM), Marker Controlled Watershed Segmentation (MCWS) and Region Growing (RG) image segmentation processes were carried out for detection of masses in mammography images [29]. The authors mentioned that the Region Growing method gives improved results when evaluated to other methods examined.

## 3. Feature Extraction

The complexity of feature removal is truely cited by Wikipedia as follows. "In all cases, removing computer quantifiable features is a skill and with the exclusion of some neural networking and genetic algorithms that mechanically perceive 'features', hand assortment of good features outlines the basis of roughly all classification algorithms". Once a figure is preprocessed to gain the improved image, it is suppled as an input to the subsequent processing stage, which can be feature mining or image categorization. The improved images are developed to remove appropriate selective features for the subsequent processing segment, which can be the image categorization segment. The importance of the mined features establishes the performance of the by and large system. In this section, authers discussed four different computational methods for feature extraction, which are statistical, shape, spectral and topological features. These features are organized on the basis of the feature classification scheme proposed by Gurevich et al. [25].

### 3.1. *Statistical Features*

The application of statistical features is speedily developing and one can observe numerous research papers illustrating the role of statistical features. In scheming statistical features, we suppose that the image examination is a recognition of a field of arbitrary numbers [30]. The most broadly utilized features of this category are histograms of intensity levels, co-occurrence matrices, entropy, fractal dimensions and statistical moments. In addition, features based on the wavelet transform can be considered as statistical features. Lakshmi et al. [26] preprocessed mammograms with the wavelet transform. The authors applied a combination of two sets of features removed from the images. Firstly, the diagonal scaling matrix 'S' (also referred as $\sum$) was acquired from the Singular Value Decomposition (SVD) of the LL band of the wavelet transform and used as one of the feature set. The second set of significant features was sliced out from a larger set of Jacobi polynomials using the Ant Colony Optimization (ACO) algorithm. The authors asserted based on their experimental outcomes that of the use of the Support Vector Machine (SVM) classifier on the picked Jacobi moments in combination with 'S' matrix resulted in a better classification rate.

Kilic et al. [27] removed features by implementing multilevel wavelet decomposition of the mammograms. The application of proper low pass filter and high pass filter to rows and columns of the images crumbled the images into four subbands. These subbands were termed the approximation, which is the small frequency constituent and detail, which is the high frequency constituent. The approximation (A) indicated a low resolution of the original image, whereas horizontal (H), vertical (V) and diagonal (D) stand for the detail coefficients. To extra classify the images, the Artificial Neural Network (ANN) classifier was utilized, in which only the estimate coefficients were presented to the ANN classifier. The authors examined that level-2 wavelet decomposition coefficients have rationally convenient number of features, which were considerably enough to create enhanced classification performance. In addition, Sakka et al. [18] applied a similar computational technique to crumble the images with focus on the high occurrence constituent. The authors found in their experimental results, and mentioned that the least asymmetric Daubechies' wavelets outcome in the incremented percentage of energy equivalent to the high frequency components as evaluated to other wavelet transformations.

Jasmine et al. [6] applied another wing of wavelet transformation to gain supplementary features from the images. The contourlet transformation decomposes images into directional subbands at several levels. Filters from two-channel non sub sampled 2-D filter bank were utilized by the authors to crumble the images. The contourlet coefficients of four sub bands were used as feature vectors to characterize an image. These features were further reduced by summing a predefined number of energy values collectively. The reduced feature set was fed to an ANN system to classify the images. Biswas et al. [28] used a model that characterised every mammogram by textural models that were characterized by the combination of Gaussians. Multiscale oriented filters were used to obtain texture maps. A generative model was suggested by the authors that could distinguish the architectural deformation in digital mammograms with the help of distinguishing textures in mammograms.

Tzikopoulos et al. [29] revealed that the determination of breast thickness and detection of breast irregularity can be efficiently used for mammogram categorization. The authors recommended a new fractal aspect related feature besides the statistical features for breast thickness classification. In all, a list of 38 such features was utilized to estimate breast density Fatty, Fatty-Glandular, Dense-Glandular and asymmetry revealing.

Mohanalin et al. [30] utilized the fuzzy entropy techniques to augment the microcalicification present in mammograms. Fuzzification of the preprocessed image offered the features that were used in the microcalcification identification stage. A Gaussian fuzzy membership function converted the preprocessed image intensity values in an interval between 0 and 1. The authors evaluated the outcomes achieved using the Tsallis Entropy and type II fuzzy against a combination of non-fuzzy and conventional Shannon entropy partition methods. Better performance was mentioned with the recommended fuzzy technique. Pradeep et al. [31] depicted how to estimate texture, statistical and structural features from a mammogram image.

### 3.2. *Shape Features*

Shape features are a set of primitives with matching associations and characteristics. Oh et al. [32] used straight forward feature mining of images as they focused on the suggestion for an efficient significance feedback system for Case Based Image Retrieval (CBIR). The d-dimensional vector that match up to the removal of primitive image features such as texture and shape was utilized to symbolize an image. A resemblance coefficient among a inquiry image and a database image was established and a technique for significance feedback was suggested. Sampaio et al. [33] suggested to take out features from the sections that might include masses via the cellular neural network method. Shape features together with eccentricity, circularity, density, circular disproportion and circular density of the section were acquired.

A well acknowledged Support Vector Machine (SVM) classifier was applied to decide whether the region of interest was masses or non-masses. Buciu et al. [34] used the Gabor wavelet filters at numerous scales to mine directional features at diverse directions and frequencies. Before the segment of feature extraction, the authors crumbled the region surrounding the irregularity into a number of patches. Principal Component Analysis (PCA) method was used for dimensionality reduction. The abridged set of features was passed to proximal support vector machines to sort the images. The authors mentioned that Gabor features were important as evaluated to original image features and gave a better classification performance. Bhattachrya et al. [40] gave importance to the precise investigation of shape features and margin of tumour mass emergeing on the breast. Fourier descriptors were used to pull out shape features. The genetic algorithm technique was used to decide a set of effective feature vectors from a huge number of feature vectors equivalent to an original mammogram. The Adaptive Neuro-Fuzzy (ANF) model was used for final classification of mammograms.

### 3.3. *Spectral Features*

The computation of spectral features is based on depiction of the illustration as a enumerated signal. Features founded on the computation of gradient should also be categorized as spectral features fairly than shape features because their verification is based on spatial frequency [30]. Tzikopoulos et al. [29] proposed the outline based level set technique for mammography mass revealing. A feature entrenched vector appreciated contour-based level set method with relaxed shape limitation was recommended. Preliminary edge on the smoothed mammogram was established using contour-based level set method.

To settle down the shape boundary, a stopping function was decided in subsequent vector-valued level set method. The attributes of the mammogram were extorted from texture maps, gradient maps and original intensity map of the image. Nine feature maps were obtained, including the texture image decomposed by morphological component analysis, gradient variations, magnitudes and orientations of gradients. The technique utilized by the authors was stated to be more efficient and strong in identifying multifarious masses when evaluated to the existing dynamic contour methods.

### 3.4. *Topological Features*

Topological or geometrical features referred to the results of applying topological functions to analyse image. These features characterize the image topology, which are continuously connected parts of an image [30]. Camilus et al. [20] extorted features with the watershed transformation technique. The original mammogram was changed into the morphological gradient image by using morphological operators. The even morphological gradient image was then utilized in a watershed study for pectoral muscle detection in the mammogram. They recommended a amalgamation algorithm to remove the over-segmentation crisis of watershed analysis. In addition, Xu et al. [19] applied a marker-controlled watershed transformation method on smooth morphological gradient image to extract features.

## 4. *Mammography* Applications

Mammograms are used as one of the most excellent existing instruments for early breast cancer revealing. Hence, the majority of the research that deals with mammograms is expressed towards an automatic finding of anomalies in mammograms. However, a few other motivating research areas are also discoverd by the researchers. In this section, authers have discussed such diverse applications including the major area of CAD. Many investigators are investigating mammograms for sevral realistic applications. One helpful application is the determination of the percentage of Glandular Tissue Composition (GTC) in Computed Radiography Mammograms (CRM) [41]. Such computation is functional in determining an evaluation of individual patient revelation dose and calculation of disease. The authors produced breast phantom images with different amalgamations of fat, glandular tissue and thickness of the entire breast. A association was established between each pixel value on CRM to the glandular tissue ratio. A reference table for conversion was formulated accordingly and the method was tested to estimate GTC. The general applications of mammography can be summarized on the basis of computer aided detection, image compression and content based image recovery.

### 4.1. *Computer Aided Detection*

Computer Aided Detection (CADe) and Computer Aided Diagnosis (CADx) with reference to breast cancer are measures in medicine that aid doctors or radiologists in the understanding of mammography images. The trials engage different stages and many researchers have put forward different methods to successfully execute definite phases. CADe systems track digitised mammography images for abnormal areas of density, mass or calcification that may indicate the incidence of cancer. The CADe system highlights these regions in the images, alerting the radiologist to carefully evaluate this area.

Digital mammography, which is also called Full Field Digital Mammography (FFDM), is a mammography system in which x-ray film is replaced by electronics that convert x-rays into mammography pictures of the breast. These systems are similar to those found in digital cameras and their efficiency enables better pictures with a lower radiation dose. The images of the breast are transferred to a computer for a radiologist to review and for long term storage. The experience of a patient during a digital mammography is similar to having a conventional screen-film mammography.

Alfonso Rojas et al. [58] reviewed the techniques employed in CAD systems. Their emphasis was to study, the model shift witnessed within the Machine Learning research community and how this has been mirrored in the design of CAD systems for breast cancer based on mammography study. They studied the ML-based CAD systems for mammography study advanced in the span of recent 10 years span. They observed that the dominance of Deep Learning in this field was not as extensive as in other ML applications. Based on the review, they concluded that the Feature Engineering and the Feature Learning methodologies, presently co-occur. One can obtain the best performance by fusion strategies that combine both. Large high-quality mammography/DBT datasets could encourage the development of Deep Learning methods in the near future, but these are not easy to assemble.

Ehsan Kozegar1 et al. [59] conducted a survey. Their focus was on 3-D automated breast ultrasound imaging modality which lead to a revolution for early detection of breast cancer. Authors studied four main components including, pre-processing, candidate regions extraction, classification, feature extraction and selection which was described for all CADe systems for automated 3-D breast ultrasound system known as ABUS images. Besides, state-of-the-art methodologies for each element were surveyed and their innovations as well as their restrictions were discussed. Few ideas to address the challenges associated to each element have been presented in this paper.

**4.1.1.** *Identification of Microcalcification*

Microcalcifications are small deposit of calcium element in the human breast that causes cancer, which can be identified from the shape and pattern of the calcium specks. They can be either benign and malignant, which cannot be felt, but detected using a mammogram. Sersic et al. [21] used the discrete wavelet transform based image enhancement method to detect microcalcification. Lakshmi et al. [26] classified mammograms as normal, abnormal, malignant and benign using the SVM machine learning classifier. A back propagation neural network classifier was proposed by Paradkar et al. [10] to detect areas of microcalcification. Jasmine et al. [6] proposed a classifier system to detect microcalcification using the combination of non sub sampled contourlet transformation and ANN. Sankar et al. [36] proposed a fractal modelling method for microalification detection. A fractal encoding method was used to model mammogram, which is visually similar to the candidate mammogram.

Buciu et al. [34] proposed the use of the proximal support vector machines to classify tumour present in mammograms. Mohanalin et al. [30] applied the Tsallis entropy and type II fuzzy indexes to automatically detect microcalcification. Sakka et al. [18] decomposed the mammograms into different frequency sub-bands to further detect the microcalcifications. Bhattacharya et al. [35] proposed Genetic algorithm (GA) based hybrid NeuroFuzzy classifier to classify masses. Sampaio et al. [33] used cellular neural network to segment the region that might contain the masses. The SVM classifier was used to determine whether the region had masses or not. The detection of mammogram masses as malign or benign was proposed by Kilic et al. [27]. The wavelet based feature vectors were supplied to ANN as inputs to determine the class of the masses. An improved edge detection algorithm was proposed by Senthilkumar et al. [6] and tested on sample of mammograms. The authors observed that the proposed edge detection method was helpful in detecting nipple point and cancer area. Thangavel et al. [37] have also presented a review of various methods proposed for the automatic detection of microcalcification in mammograms.

Most of the information is hidden to the human observer, in the original mammography images obtained by X-ray radiography. Redundant discrete wavelet transform method, which is shift invariance and numerically robust was presented by Seršice and Lonèariœ [16]. The procedure consists of three steps, which are low-frequency tissue density component removal, noise filtering and microcalcification enhancement. The method had been applied to a number of mammogram images to show promising results.

Sakka et al. [18] presented a comparative study of the orthogonal wavelets using the evaluation criterion of the degree of similarity between the wavelet and image profile. This guides the decision on the appropriate family of wavelets to use for mammography image processing. Bazzani et al. [38] used the wavelet transform to detect structures having smaller sizes than 1mm by performing analysis of multiresolution. Ferreira et al. [39] applied the wavelet transformation using the Daubechies 4 and Haar wavelets in the image decomposition process. They kept only the coefficients with the largest in magnitude of the decomposed image in the first level of decomposition.

Sentelle et al. [40] employed the wavelet analysis method to detect calcifications in mammograms. The wavelet was processed by simply removing the lowest resolution approximation coefficients. This was followed by performing an inverse wavelet transform whose output was appropriately thresholded to provide a binary detection of calcifications. Soltanian-Zadeh et al. [41] used the Daubechies 6, 10 and 12 wavelets in a feature extraction method based on image decomposition. The calculation of entropy and energy in each of the sub-bands was then performed. They also applied the multi-wavelet transformations with GHM, CL and SA4 multi-wavelets using several scaling functions and mother wavelets.

Lambrou et al. [42] used the Daubechies 4-TAP wavelet filter in all wavelet architectures. They collected the first and second order statistical values with grey level run length measurements from all the signals and corresponding wavelet coefficients. Yoshida et al. [43], [44] applied the wavelet transformation using the least asymmetric Daubechies wavelets with length 8, 12 and 20 to enhance microcalcifications detection. The Haar and Daubechies 4 wavelets were used by Ferreira et al. [39] to evaluate a supervised learning classifier by transforming images in a wavelet basis. In the literature [45, 46], the ability to visualise microcalcification using the state of the art sonographic equipment has been described. The ability of sonography to depict mammographyally microcalcification is further described in the reports [47, 49, 50]. Moon et al. [48] reported that with sonography, 77% of DCIS cases were visible as a breast mass associated with microcalcification. They reported that calcifications associated with malignant tumours were more likely to be seen on sonography.

**4.1.2.** *Lesion Segmentation*

Lesion segmentation is one of the critical steps of CAD as a result, several researchers have proposed novel methods for segmenting lesions from mammograms. Effective lesion segmentation will ultimately contribute to the overall performance of automatic detection of microcalcification or masses. Xu et al. [19] enhanced the conventional watershed method for reliable segmentation of lesions using internal and external markers. Wang et al.[51] used a feature embedded vector valued contour based level set method with a relaxed shape constraint algorithm to effectively find mass regions in mammograms.

The work of Tzikopoulos et al. [29] focused on mammography segmentation and classification. Rahmati et al. [4] proposed the Fuzzy contrast limited adaptive histogram equalization filter to attain increased segmentation accuracy. Nagi et al. [11] used the morphological preprocessing followed by a seeded region growing algorithm to detect region of interest.

### 4.1.3. Pectoral Muscle Identification

Pectoral muscle identification can be considered as a preprocessing step in the complete process of CAD. In particular, amongst the various phases of CAD, Camilus et al. [20] and Bandyopadhyay et al. [24] proposed methods for pectoral muscle identification and suppression. Specifically, Camilus et al. [20] applied the watershed transformation method to identify the pectoral muscle. In the case of Bandyopadhyay et al. [24], they used a simple algorithm to determine a region that covers the entire area of pectoral muscle. Moreover, Liu et al. [12] improved the traditional gradient Vector Flow Snake (GVFS) algorithm to extrapolate breast region. In the first phase, rough breast border was obtained by the morphological erosion method. The modified edge map based on the gradient adjustment and GVFS was used to obtain an accurate breast border from the rough breast border. The anatomically oriented breast coordinate system was proposed for mammogram analysis by Brandt et al. [52]. The authors demonstrated that the proposed coordinate transformation method can be used to extract Gaussian derivative features without nonlinearly deforming the images. Architectural distortion on mammogram can be recognized by the method proposed by Biswas et al [28], which was based on multiscale texture modelling method.

### 4.2. *Image Compression*

Mammograms are generally high resolution and large size images, which require the need for compression to ease transportation across the low bandwidth computer networks. Researchers are exploring various methods to compress mammograms without distorting the subtle information contained with the images. AbuBaker et al. [53] proposed a preprocessing method for reducing the size and enhancing the quality of mammograms. The pixel depth conversion algorithm was used as a mammogram shrinking method that effectively resulted in the average reduction of 87% in size with no loss of data in the breast region. Tan et al. [54] proposed to train auto encoders using the image patches of an original mammogram instead of a whole mammogram. The mammograms with large sizes resulted in difficulty to train the autoencoders. The comparison between the compression performance of different types of autoencoders with the proposed method was presented. The effect of using different sizes of image patches was also presented.

**4.3.** *Content Based Image Retrieval*

Content based image retrieval (CBIR) engages recovery of pertinent or alike images corresponding to a query image. It is a next step of image recognition. Particularly in bioscience field the success to CBIR has been quite limited. Oh et al. [32] proposed an adaptive CBIR system that used an advanced relevance feedback method for incremental learning. Wei et al. [55] used a CBIR method retrieve matching mammograms with a query image from the database to the user. The authors had interpreted the mammography lesions on the basis of their medical distinctiveness particularly in the Breast Imaging Reporting and Data System (BI-RADS) principles. With the help of proposed hierarchical similarity measurement, the features of query image were matched with other images and an appropriate relevance feedback mechanism was also proposed to improve the retrieval performance.

**5. Screen-film and Digital Mammograms**

The Screen-Film Mammogram (SFM) is created on film and a radiologist has to inspect the generated image on a beam box. This tool has been applied in identifying breast cancers, but it is not ideal. For instance, in India, the incidence of breast cancer is rapidly increasing with an estimated 80,000 new cases diagnosed annually [62]. The survey illustrates that 92% of breast imaging clinics India necessitate digital mammogram, but are using SFM [63]. The use of SFM, which is also called the analogue mammography has been widely exploited. It is still believed to be extremely superior at detecting breast irregularities at an early phase. In addition, the SFM is widely used by most health insurance providers because of following features of SFM:

Digital mammography has been made available in many major cities, but may not be ubiquitous because it is cost expensive to procure. Digital mammography systems cost about 1.5 to 4 times more than the screen-film systems. The two methods, have a similar level of accuracy of 92 % at ruling out breast cancer and percentage of false positive is the same for both methods [64]. Some radiology centres say that digital equipment delivers less radiation, but most experts note that the level of radiation in screen-film mammography is extremely low [65]. Although digital mammography has many potential advantages over traditional screen-film mammography, clinical trials have shown that the overall diagnostic accuracy levels of the current digital mammography and screen-film mammography are similar when used in breast cancer screening [65]. Robert D. Rosenberg et al. [80] conducted study and their findings indicates the range of performance standards for screening mammography performed by group of experts. Outcome of the study is useful as proportional data for particular radiologists and for formation of consequent strategies.

Joshi et al. [63] discussed various application areas of mammograms to include CAD, image compression and Content Based Image Retrieval (CBIR). Berns et al. [64], using the screen-film mammography and soft-copy digital mammography compared the acquisition time, interpretation time. They observed that the use of digital mammography for screening examinations significantly shortened acquisition time, but significantly increased interpretation time. In other related studies, comparisons between the screen-film mammography and digital mammography were done [69, 70, 71, 72]. It was found that almost all (91%) digital mammography facilities in the United States use computer aided detection when compared to the 49% users of screen-film screening facilities [71]. Kerlikowske [68] showed evidence that the merits should outweigh the demerits, before the medical community adopts new technologies, specifically regarding computer-aided detection.

Iared et al. [67] focused on comparing the performance of digital mammography with screen-film mammography in terms of cancer detection rates, patient recall rates and characteristics of the detected tumours. Kerlikowske et al. [65] found that the proportion of cancer cases diagnosed at an early stage varies as a result of sensitivity and specificity of each modality by age, tumour characteristics, breast density and menopausal status. The studies to date do not support the apparent merits of digital mammography over screen-film mammography when used for diagnostic purpose [72, 73, 74, 75]. This implies more practical research effort on improving the technology of digital mammography. Bhale et al. [71] compared the features of digital mammograms with those of screen-film mammograms in order to analyse their significance. The various image enhancement methods were applied on the screen-film mammograms and evaluated using the methods of Peak Signal to Noise Ratio (PSNR) and Root Mean Square Error (RMSE). The result showed that the clarity of processed screen-film mammograms is as good as digital mammograms.

Koomen et al. [73] discussed the use of newer versions of mammography, such as digital mammography with tomosynthesis and digital subtraction mammography with different tools for women at high risk. The emphasized the use of tools that might be useful for less invasive therapy of breast cancer with imaging to monitor therapy efficacy. Bhale et al. [71] presented a comparative analysis of digital mammography and screen-film mammography. After the implementations of preprocessing methods, they performed statistical analysis using PSNR and RMSE methods and they compared the results with the Mean Opinion score (MOS). In addition, they analysed the extent of radiation exposed during digital mammography and screen-film mammography. They found that the cost factor is higher in the case of digital mammography, but quality is good. They conclude that the cost of screen-film mammography is less, but quality can be enhanced by applying image enhancement methods to preprocess the images.

## 6. Development of Mammogram Corpora

The online community of practice where access to data is made open and simple as well as allow for sharing of research results in a real time can be an effective method for robust screening of mammograms. Digital corpora for screening mammograms are an important resource within such a community for collaborating researchers to study digital mammograms through computational methods. Mammogram corpora focus on the context of mammogram analysis to assist experts in the screening process of breast cancer. The corpora should intrinsically contain a substantial volume of cases of normal breast and cancer breast. In general, obtaining reliable medical data for experiments is crucial for effective diagnosis and treatment of diseases. Consequently, researchers are working on the development mammogram corpora to aid the process of mammogram detection and diagnosis. Youn et al. [57] described a procedure to create a realistic database of digital mammograms, wherein they used different modelling methods to generate noise-free mammograms.

The use of benchmark corpora to evaluate the performance of computational methods is generally one of the essential steps in contemporary research. Several standard corpora are being continuously developed for different research areas. The availability and use of standard corpora are rapidly increasing, making the development and management of corpora an important research topic. However, in the particular instance of Medical Image Processing (MIP) domain, the process of collecting data from patients presents enormous challenges. The challenges are caused by inhibiting factors such as willingness of patients to share information, large size of images, computational capabilities of experts and multi-modality of information. All of these factors play an important role in developing standard corpora in the area of MIP and particularly for breast cancer related mammogram images. Hence, relatively few standard corpora are available in the MIP area to support researchers with reliable data for experimental works.

Suckling et al. [82] elaborated on the process of developing the Mammograhy Image analysis Society (MIAS) corpus. The corpus represented a constructive and practical contribution to the research in the MIP, particularly in the mammography research. Here are a few references that illustrates the use of standard corpora in different time spans from 1994 to 2016 in the MIP field. Mudigonda [81] in 2001 used the mini MIAS [78, 81] corpus of 56 images. The size of each image in the corpus is 1024 X 1024 pixels at a resolution of 200 meters, including 30 benign breast masses, 13 malignant and 13 normal cases. They proposed a method for the detection of masses in mammography images that employs the Gaussian smoothing and sub sampling operations as preprocessing steps. They used the MIAS corpus to test the performance of their method. The mass regions were successfully segmented to classify benign or malignant disease using five texture features based on Gray level Co-occurrence Matrices (GCMs) and features in a logistic regression method.

Rangaraj M. Rangayyan [83] in 2006 used 19 mammograms exhibiting architectural distortion, from the Mini-MIAS corpus [78]. They presented a new method to detect and localise architectural distortion by analysing oriented texture in mammograms. A bank of Gabor filters was used to obtain the orientation field of the given mammogram. Rangaraj M. Rangayyan [83] in 2010 obtained mammograms from the corpus of 1,745 digitised mammograms of 170 subjects from the screen test of the Alberta programme for the early detection of breast cancer [84, 85]. They performed operations of CAD, using this corpus for the architectural detection of mammograms. Matheus Bruno Roberto Nepomuceno et al. [86] presented mammogram corpus that was judged to provide sufficient number of images of high quality with an inbuilt functional search system in comparison with other corpora.

The MIAS, MINI-MIAS (In mini MIAS database the original MIAS Database (digitised at 50 micron pixel edge) has been reduced to 200 micron pixel edge and clipped/padded so that every image is 1024 × 1024 pixels) and Digital Database for Screening Mammography (DDSM) corpora are available on the public internet. These corpora follow certain benchmark characteristics such as database reference number, character of background tissue (fatty, fatty glandular, dense glandular), class of abnormality present in a mammogram and severity of the abnormality [85]. These corpora acquire features like mammograms in different angles like Right Medio Lateral Oblique, Left Medio Lateral Oblique. They labelled each image with the date of study, age of a patient and type of mammogram. Bhale et al. [78] discussed the procedure for collecting mammograms of different modalities. They presented details of methods used to reduce the size of image and enhance image clarity. Their corpus acquires important features such as the SFM and DM mammograms along with the mammogram types. The acquired information also includes, clinical history of a patient such as age, number of kids, dietary habits, canner family history and histopathology reports.

The practical application of digital mammography research is generally challenging with a recent open project tagging it as a Digital Mammography DREAM Challenge (DM Challenge). The DM Challenge poses fundamental questions about systems biology and translational medicine. It is one of two large prize Coding4Cancer challenges and amongst the several open grand challenges in the field of Biomedical image analysis published in 2016 (http://grand-challenge.org/all_challenges/). The DM Challenge corporal promises to host data consisting of more than 641,000 digital mammograms with the corresponding patient characteristics and outcome measures (https://www.synapse.org/#!Synapse:syn4224222/wiki/401743). The goal of the DM Challenge, according to the source, is to apply an open science, deep learning and crowd sourced methods to develop algorithms for risk stratification of screening mammograms to help improve breast cancer detection. These algorithms as declared by the source, should potentially benefit the interpretation of other tumour imaging and impact a wide group of cancer patients. The website of the DM Challenge, which is opened to the desiring participants, is designed and run by a community of researchers from a variety of organizations. The expertise and institutional support of the website are provided by the Sage Bionetworks along with the infrastructure to host challenges through their Synapse platform. Figure 2 is an example of the interface to the website of the DM challenge for researchers willing to participate.

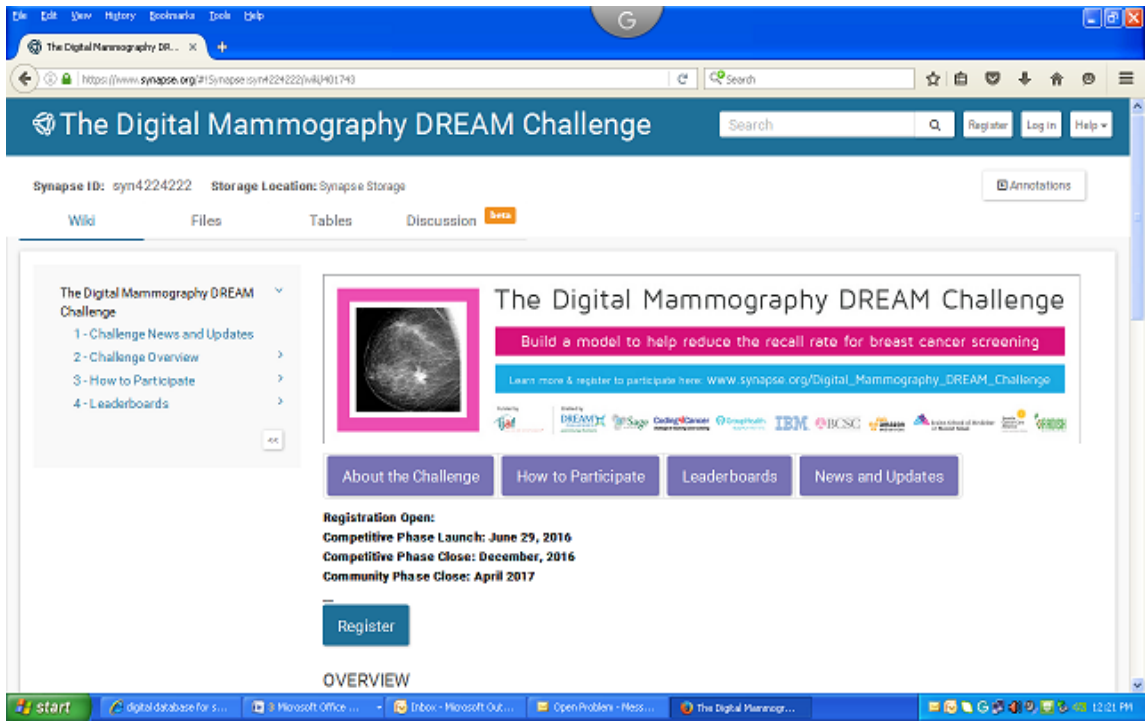

Figure 2 A Web Interface of the Digital Mammography DREAM Challenge. Sourced from https://www.synapse.org/#!Synapse:syn4224222/wiki/401743.

## 7. Conclusion

The application of intensive computational methods such as open science, deep learning, crowd sourced and digital image processing is generally believed to improve breast cancer detection and diagnosis. The algorithms based on these computational methods should potentially benefit effective interpretation of tumour imaging and impact positively on a wide group of cancer patients. The perfect analysis of raw mammograms could not be achieved by radiologists without enhancement tools because of the presence of artefacts and ambiguity in mammograms. The unfoldment of mammograms using computational analysis methods to obtain significant features is possible because of the ongoing research in digital mammography. In particular, researchers have revealed that image processing methods and tools could be suitable to discover significant hidden facts in mammograms.

This paper discussed recent studies that have contributed to the application of computational methods for automating mammograms. The various methods for image preprocessing, feature extraction, mammography applications, screen-film versus digital mammograms and development of mammogram corpora were conspicuously discussed. The information provided in this paper will undoubtedly empower and enrich the knowledge of young researchers, experienced researchers and medical practitioners. Moreover, it will serve as a roadmap or directional compass to improve the state of the art computational methods in mammography research and related studies.

**References**

1. Sivaramakrishna, R., Obuchowski, N. A., Chilcote, W. A., Cardenosa, G. and Powell, K. A., 2000. Comparing the performance of mammographic enhancement algorithms: a preference study. *American Journal of Roentgenology*, *175*(1): 45-51.
2. Pizer, S. M., Amburn, E. P., Austin, J. D., Cromartie, R., Geselowitz, A., Greer, T., Romeny, B. H., Zimmerman, J. B. and Zuiderveld, K., 1987. Adaptive histogram equalization and its variations. *Computer Vision, Graphics, and Image Processing*, *39*(3): 355-368.
3. Senthilkumar, B. and Umamaheswari, G., 2011. A novel edge detection algorithm for the detection of breast cancer. *European Journal of Scientific Research*, *53*(1): 51-55.
4. Rahmati, P., Hamarneh, G., Nussbaum, D. and Adler, A., 2010. A new preprocessing filter for digital mammograms. In: *International Conference on Image and Signal Processing*, Springer Berlin Heidelberg, 6134, 585-592.
5. Sundaram, M., Ramar, K., Arumugam, N. and Prabin, G., 2011. Histogram modified local contrast enhancement for mammogram images. *Applied soft computing*, *11*(8): 5809-5816.
6. Jasmine, J. L., Govardhan, A. and Baskaran, S., 2010. Classification of microcalcification in mammograms using nonsubsampled contourlet transform and neural network. *European Journal of Scientific Research*, 531-539.
7. Abraham, N. P. and Ponraj, N., 2012. Segmentation methods for locating masses and locating masses and locating breast boundaries: A survey. *International Journal of Advanced Research in IT and Engineering*. ISSN: 2278-6244, 1(5): 9-19.
8. Ji, T. L., Sundareshan, M. K. and Roehrig, H., 1994. Adaptive image contrast enhancement based on human visual properties. *IEEE Transactions on Medical Imaging*, *13*(4): 573-586.

9. Panetta, K., Zhou, Y., Agaian, S. and Jia, H., 2011. Nonlinear unsharp masking for mammogram enhancement. *IEEE Transactions on Information Technology in Biomedicine*, *15*(6): 918-928.

10. Paradkar, S. and Pande, S. S., 2011. Intelligent detection of microcalcification from digitized mammograms. *Sadhana*, *36*(1): 125-139.

11. Nagi, J., Kareem, S. A., Nagi, F. and Ahmed, S. K., 2010. Automated breast profile segmentation for ROI detection using digital mammograms. In: *Biomedical Engineering and Sciences, IEEE EMBS Conference,* 87-92.

12. Liu, C. C., Tsai, C. Y., Tsui, T. S. and Yu, S. S., 2012. An improved GVF snake based breast region extrapolation scheme for digital mammograms. *Expert Systems with Applications*, *39*(4): 4505-4510.

13. Morrow, W. M., Paranjape, R. B., Rangayyan, R. M. and Desautels, J. E. L., 1992. Region-Based contrast enhancement of mammograms. *IEEE transactions on Medical* Nordin *Imaging*, *11*(3): 392-406.

14. Nordin, Z. M., Isa, N. A. M., Ngah, U. K. and Zamli, K. Z., 2008. Mammographic Image contrast enhancement through the use of moving contrast sweep. In: *International Conference on Knowledge-Based and Intelligent Information and Engineering Systems,* Springer Berlin Heidelberg, 533-540.

15. Rangayyan, R. M., Ayres, F. J. and Desautels, J. L., 2007. A review of computer-aided diagnosis of breast cancer: Toward the detection of subtle signs. *Journal of the Franklin Institute*, *344*(3-4): 312-348.

16. Seršice, D. and Lonèariœ, S., 1998. Enhancement of mammographic images for detection of microcalcifications. In: *Signal Processing Conference (EUSIPCO), 9th European,* IEEE, 1-5.

17. https://www.breastcancer.org/symptoms/understand_bc/statistics

18. Sakka, E., Prentza, A., Lamprinos, I. E. and Koutsouris, D., 2006. Microcalcification detection using multiresolution analysis based on wavelet transform. In: *International Special Topic Conference on Information Technology in Biomedicine, Ioannina, Epirus, IEEE*, 32, 33.

19. Xu, S., Liu, H. and Song, E., 2011. Marker-controlled watershed for lesion segmentation in mammograms. *Journal of Digital Imaging*, *24*(5): 754-763. Xu
20. Camilus, K. S., Govindan, V. K. and Sathidevi, P.S., 2011. Pectoral muscle identification in mammograms. *Journal of Applied Clinical Medical Physics*, *12*(3): 215-230.
21. Damir sersic and Seven Loncaric, et al "Enhancement of mammographic images for detection of microcalcifications", Department of Electronic Systems and Information Processing, Faculty of Electrical Engineering and Computing, University of Zagreb, Vukovar Avenue 39, CROATIA
22. Ponraj, D. N., Jenifer, M. E., Poongodi, P. and Manoharan, J. S., 2011. A survey on the preprocessing techniques of mammogram for the detection of breast cancer. *Journal of Emerging Trends in Computing and Information Sciences*, *2*(12): 656-664.
23. Moradmand, H., Setayeshi, S. and Targhi, H. K., 2012. Comparing methods for segmentation of microcalcification clusters in digitized mammograms. *arXiv: 1201.5938, 8(6)*.
24. Bandyopadhyay, S. K., 2010. Pre-processing of mammogram images. *International Journal of Engineering Science and Technology*, *2*(11): 6753-6758.
25. Gurevich, I. B. and Koryabkina, I. V., 2006. Comparative analysis and classification of features for image models. *Pattern Recognition and Image Analysis*, *16*(3): 265-297.
26. Lakshmi, N. S. R. and Manoharan, C., 2011. A novel hybrid ACO based classifier model for mammogram microcalcifications using combined feature set with SVM. *European Journal of Scientific Research*, *53*(2): 239-248.
27. Kilic, N., Gorgel, P., Ucan, O. N. and Sertbas, A., 2010. Mammographic mass detection using wavelets as input to neural networks. *Journal of medical systems*, *34*(6): 1083-1088.
28. Biswas, S. K. and Mukherjee, D. P., 2011. Recognizing architectural distortion in mammogram: a multiscale texture modeling approach with GMM. *IEEE Transactions on Biomedical Engineering*, *58*(7): 2023-2030.

29. Tzikopoulos, S. D., Mavroforakis, M. E., Georgiou, H. V., Dimitropoulos, N. and Theodoridis, S., 2011. A fully automated scheme for mammographic segmentation and classification based on breast density and asymmetry. *Computer Methods and Programs in Biomedicine*, *102*(1): 47-63.

30. Mohanalin, Beenamol, Kalra, P. K. and Kumar, N., 2010. A novel automatic microcalcification detection technique using Tsallis entropy & a type II fuzzy index. *Computers & Mathematics with Applications*, *60*(8): 2426-2432.

31. Pradeep, N., Girisha, H., Sreepathi, B. and Karibasappa, K., 2012. Feature extraction of mammograms. *International Journal of Bioinformatics Research*, *4*(1): 241-244.

32. Oh, J. H., Yang, Y. and El Naqa, I., 2010. Adaptive learning for relevance feedback: application to digital mammography. *Medical Physics*, *37*(8): 4432-4444.

33. Sampaio, W. B., Diniz, E. M., Silva, A. C., De Paiva, A. C. and Gattass, M., 2011. Detection of masses in mammogram images using CNN, geostatistic functions and SVM. *Computers in Biology and Medicine*, *41*(8): 653-664.

34. Buciu, I. and Gacsadi, A., 2011. Directional features for automatic tumor classification of mammogram images. *Biomedical Signal Processing and Control*, *6*(4): 370-378.

35. Bhattacharya, M., Sharma, N., Goyal, V., Bhatia, S. and Das, A., 2011. A study on genetic algorithm based hybrid soft computing model for benignancy/malignancy detection of masses using digital mammogram. *International Journal of Computational Intelligence and Applications*, *10*(02): 141-165.

36. Sankar, D. and Thomas, T., 2010. A new fast fractal modeling approach for the detection of microcalcifications in mammograms. *Journal of Digital Imaging*, *23*(5): 538-546.

37. Thangavel, K., Karnan, M., Sivakumar, R. and Mohideen, A. K., 2005. Automatic detection of microcalcification in mammograms–A review. *International Journal on Graphics Vision and Image Processing*, *5*(5): 31-61.

38. Bazzani, A., Bevilacqua, A., Bollini, D., Campanini, R., Lanconelli, N., Riccardi, A. and Romani, D., 2000. Automatic detection of clustered microcalcifications using a combined method and an SVM classifier. In: 5[th] *International Workshop on Digital Mammography*, 161-167.

39. Ferreira, C. B. R. and Borges, D. L., 2003. Analysis of mammogram classification using a wavelet transform decomposition. *Pattern Recognition Letters*, *24*(7): 973-982.
40. Sentelle, S., Sentelle, C. and Sutton, M. A., 2002. Multiresolution-based segmentation of calcifications for the early detection of breast cancer. *Real-Time Imaging*, *8*(3): 237-252.
41. Soltanian-Zadeh, H. and Rafiee-Rad, F. and Pourabdollah-Nejad D. S., 2004. Comparison of multiwavelet, wavelet, Haralick, and shape features for microcalcification classification in mammograms. *Pattern recognition*, *37*(10): 1973-1986.
42. Lambrou, T., Linney, A. D., Speller, R. and Todd-Pokropek, A., 2002. Statistical classification of digital mammograms using features from the spatial and wavelet domains. *Medical Image Understanding and Analysis*, 22-23.
43. Yoshida, H., Doi, K., Nishikawa, R. M., Muto, K. and Tsuda, M., 1994. Application of the wavelet transform to automated detection of clustered microcalcifications in digital mammograms. *Academic Reports of Tokyo Institute of Polytechnics*, *16*, 24-37.
44. Yoshida, H., Zhang, W., Cai, W., Doi, K., Nishikawa, R. M. and Giger, M. L., 1995. Optimizing wavelet transform based on supervised learning for detection of microcalcifications in digital mammograms. In: *International Conference on Image Processing,* IEEE*,* 3, 152-155.
45. Stavros, A. T., Thickman, D., Rapp, C. L., Dennis, M. A., Parker, S. H. and Sisney, G. A., 1995. Solid breast nodules: use of sonography to distinguish between benign and malignant lesions. *Radiology*, *196*(1): 123-134.
46. Kasumi, F., 1988. Can microcalcifications located within breast carcinomas be detected by ultrasound imaging? *Ultrasound in Medicine & biology*, *14(1)*: 175-182.
47. Yang, W. T., Suen, M., Ahuja, A. and Metreweli, C., 1997. In vivo demonstration of microcalcification in breast cancer using high resolution ultrasound. *The British Journal of Radiology*, *70*(835): 685-690.
48. Moon, W. K., Myung, J. S., Lee, Y. J., Park, I. A., Noh, D. Y. and Im, J. G., 2002. US of ductal carcinoma in situ 1. *Radiographics*, *22*(2): 269-281.

49. Hashimoto, B. E., Kramer, D. J. and Picozzi, V. J., 2001. High detection rate of breast ductal carcinoma in situ calcifications on mammographically directed high-resolution sonography. *Journal of Ultrasound in Medicine*, *20*(5): 501-508.

50. Ranieri, E., D'Andrea, M. R., D'Alessio, A., Bergomi, S., Caprio, G., Calabrese, G. B. and Virno, F., 1996. Ultrasound in the detection of breast cancer associated with isolated clustered microcalcifications, mammographically identified. *Anticancer research*, *17*(4A): 2831-2835.

51. Wang, Y., Tao, D., Gao, X., Li, X. and Wang, B., 2011. Mammographic mass segmentation: embedding multiple features in vector-valued level set in ambiguous regions. *Pattern Recognition*, *44*(9): 1903-1915.

52. Brandt, S. S., Karemore, G., Karssemeijer, N. and Nielsen, M., 2011. An anatomically oriented breast coordinate system for mammogram analysis. *IEEE Transactions on Medical Imaging*, *30*(10): 1841-1851.

53. AbuBaker, A. A., Qahwaji, R. S., Aqel, M. J., Al-Osta, H. and Saleh, M. H., 2006. Efficient pre-processing of USF and MIAS mammogram images. *Journal of Computer Science*, 3(2): 67–75.

54. Tan, C. C. and Eswaran, C., 2011. Using autoencoders for mammogram compression. *Journal of medical systems*, *35*(1): 49-58.

55. Wei, C. H., Li, Y. and Huang, P. J., 2011. Mammogram retrieval through machine learning within BI-RADS standards. *Journal of Biomedical Informatics*, *44*(4): 607-614.

56. https://cytecare.com/blog/statistics-of-breast-cancer/

57. Youn, H., Han, J. C., Cho, M. K., Jang, S. Y., Kim, H. K., Kim, J. H., Tanguay, J. and Cunningham, I. A., 2011. Numerical generation of digital mammograms considering imaging characteristics of an imager. *Nuclear Instruments and Methods in Physics Research Section A: Accelerators, Spectrometers, Detectors and Associated Equipment*, *652*(1): 810-814.

58. Domınguez, Alfonso Rojas, et al. "CAD of Breast Cancer: A Decade-Long Review of Techniques for Mammography Analysis." *Advances in Artificial Intelligence*: 115.

59. Kozegar, Ehsan, et al. "Computer aided detection in automated 3-D breast ultrasound images: a survey." Artificial Intelligence Review (2019): 1-23, Springer publications.

60. Abdelhafiz, Dina, et al. "Convolutional Neural Network for Automated Mass Segmentation in Mammography." 2018 IEEE 8th International Conference on Computational Advances in Bio and Medical Sciences (ICCABS). IEEE, 2018.
61. Ronneberger, P. Fischer, and T. Brox, "U-net: Convolutional networks for biomedical image segmentation," in International Conference on Medical image computing and computer-assisted intervention, pp. 234– 241, Springer, 2015.
62. Z. Hu, J. Tang, Z. Wang, K. Zhang, L. Zhang, and Q. Sun, "Deep learning for image-based cancer detection and diagnosis survey," Pattern Recognition, 2018.
63. Joshi, M. R. and Bhale, A. K., 2012. Computational unfoldment of mammograms. In: *Pattern Recognition, Informatics and Medical Engineering (PRIME), International Conference, IEEE,* 324-330.
64. Berns, E. A., Hendrick, R. E., Solari, M., Barke, L., Reddy, D., Wolfman, J., Segal, L., DeLeon, P., Benjamin, S. and Willis, L., 2006. Digital and screen-film mammography: comparison of image acquisition and interpretation times. *American Journal of Roentgenology*, *187*(1): 38-41.
65. Kerlikowske, K., Hubbard, R. A., Miglioretti, D. L., Geller, B. M., Yankaskas, B. C., Lehman, C. D., Taplin, S. H. and Sickles, E. A., 2011. Comparative effectiveness of digital versus film-screen mammography in community practice in the United States: a cohort study. *Annals of Internal Medicine*, *155*(8): 493-502.
66. John D. Keen, MD, MBA Stroger Hospital of Cook County Chicago, IL 60612, Annals of Internal Medicine 248 © 2012 American College of Physicians.
67. Iared, W., Shigueoka, D. C., Torloni, M. R., Velloni, F. G., Ajzen, S. A., Atallah, Á. N. and Valente, O., 2011. Comparative evaluation of digital mammography and film mammography: systematic review and meta-analysis. *Sao Paulo Medical Journal*, *129*(4): 250-260.
68. Kerlikowske, K., 2010. A call for evidence of benefits outweighing harms before implementing new technologies: comment on "Diffusion of Computer-Aided Mammography after Mandated Medicare Coverage". *Archives of internal medicine*, *170*(11): 990-991.
69. Kim, H. H., Pisano, E. D., Cole, E. B., Jiroutek, M. R., Muller, K. E., Zheng, Y., Kuzmiak, C. M. and Koomen, M. A., 2006. Comparison of calcification specificity in digital mammography using soft-copy display versus screen-film mammography. *American Journal of Roentgenology*, *187*(1): 47-50.

70. Tice, J. A. and Feldman, M. D., 2008. Full-field digital mammography compared with screen-film mammography in the detection of breast cancer: rays of light through DMIST or more fog? *Breast Cancer Research and Treatment*, *107*(2): 157-165.

71. Bhale, A. and Joshi, M., 2013. Enhancement of screen film mammogram up to a level of digital mammogram. In: *Intelligent Interactive Technologies and Multimedia,* Springer Berlin Heidelberg*, 276, 133*-142.

72. Dehghani, S. and Dezfooli, M. A., 2011. A method for improve preprocessing images mammography. *International Journal of Information and Education Technology*, *1*(1): 90-93.

73. Koomen, M., Pisano, E. D., Kuzmiak, C., Pavic, D. and McLelland, R., 2005. Future directions in breast imaging. *Journal of Clinical Oncology*, *23*(8): 1674-1677.

74. Hassanien, A. E. and Badr, A., 2003. A comparative study on digital mamography enhancement algorithms based on fuzzy theory. *Studies in Informatics and Control*, *12*(1): 21-32.

75. Adam, A. and Omar, K., 2006. Computerized breast cancer diagnosis with Genetic Algorithm and Neural Network. In: *3rd International Conference on Artificial Intelligence and Engineering Technology (ICAIET)*, 22-24.

76. Pechenizkiy, M., 2005. The impact of feature extraction on the performance of a classifier: kNN, Naïve Bayes and C4. 5. In Conference of the Canadian Society for Computational Studies of Intelligence (pp. 268-279). Springer Berlin Heidelberg.

77. Potente, G., Messineo, D., Maggi, C. and Savelli, S., 2009. Practical application of contrast-enhanced magnetic resonance mammography [CE-MRM] by an algorithm combining morphological and enhancement patterns. Computerized Medical Imaging and Graphics, 33(2), pp.83-90.

78. Bhale, A. Joshi, M. et al. 2013. Development of a Standard Multimodal Mammographic Dataset. In: Proceedings of National Conference on Advances in Computing (NCAC), North Maharashtra, University, Jalgaon, Maharashtra, India. ISBN: 978-81-910591-7-5.

79. http://www.thehindu.com/sci-tech/health/breast-cancer-a-wake-up-call-for-indian-women/article26847.ece
80. Rosenberg, R. D., Yankaskas, B. C., Abraham, L. A., Sickles, E. A., Lehman, C. D., Geller, B. M., Carney, P. A., Kerlikowske, K., Buist, D. S., Weaver, D. L. and Barlow, W. E., 2006. Performance benchmarks for screening mammography 1. *Radiology*, *241*(1): 55-66.
81. Mudigonda, N. R., Rangayyan, R. M. and Desautels, J. L., 2001. Detection of breast masses in mammograms by density slicing and texture flow-field analysis. *IEEE Transactions on Medical Imaging*, *20*(12): 1215-1227.
82. Suckling, J., Parker, J., Dance, D. R, Astley, S., Hutt, I., Boggis, C. R. M, Ricketts, I., Stamatakis, E., Cerneaz, N., Kok, S. L and Taylor, P., 1994. The mammographic image analysis society digital mammogram database. In: *Exerpta Medica, International Congress Series*, 1069, 375-378.
83. Rangayyan, R. M. and Ayres, F. J., 2006. Gabor filters and phase portraits for the detection of architectural distortion in mammograms. *Medical and Biological Engineering and Computing*, *44*(10): 883-894.
84. Rangayyan, R. M., Banik, S. and Desautels, J. L., 2010. Computer-aided detection of architectural distortion in prior mammograms of interval cancer. *Journal of Digital Imaging*, *23*(5): 611-631.
85. Matheus, B. R. N. and Schiabel, H., 2011. Online mammographic images database for development and comparison of CAD schemes. *Journal of digital imaging*, *24*(3): 500-506.